\documentclass[aip,jap,reprint,superscriptaddress,showpacs]{revtex4-1}

\usepackage{amsmath,amssymb}
\usepackage{graphicx}
\usepackage{color}
\usepackage{epstopdf}
\usepackage{multirow}

\renewcommand{\d}{{\rm d}}
\newcommand{\e}{{\rm e}}
\newcommand{\imai}{{\rm i}}

\begin{document}

\title{Simulating terahertz quantum cascade lasers: Trends from samples from different labs}
\author{David O. Winge}
\email[]{Electronic mail: David.Winge@teorfys.lu.se}
\author{Martin Francki\'e}
\author{Andreas Wacker}
\date{\today}
\affiliation{Mathematical Physics, Lund University, Box 118, 22100 Lund, Sweden}

\begin{abstract}
We present a systematic comparison of the results from our non-equilibrium Green's function formalism with a large number of 
AlGaAs-GaAs terahertz quantum cascade lasers previously published in the literature. 
Employing identical material and simulation parameters for all samples, we observe that discrepancies between measured and calculated peak currents are similar for samples from a given group. This suggests that the differences between experiment and theory are partly due to a lacking reproducibility for devices fabricated at different laboratories. Varying the interface roughness height  for different devices, we find that the peak current under lasing operation hardly changes, so that differences in interface quality appear not to be the sole reason for the lacking reproducibility. 
\end{abstract}


%72.10.-d 	Theory of electronic transport; scattering mechanisms
%72.20.-i 	Conductivity phenomena in semiconductors and insulators
%72.20.Ht 	High-field and nonlinear effects
%72.80.-r 	Conductivity of specific materials
%72.80.Ey 	III-V and II-VI semiconductors
 
\maketitle

\section{Introduction}
Since the first realization of the terahertz (THz) quantum cascade laser
\cite{KohlerNature2002} (QCL) it has now been shown to be a reliable source of
terahertz radiation, although at low
temperatures\cite{WilliamsNatPhotonics2007}. Different designs have been
proposed and  fabricated at many different laboratories of the
world. Simultaneously, simulations have been performed for a large variety of
samples with different models. These can be based on rate equations for the
electron densities  \cite{DonovanAPL1999,IndjinJAP2002},   Monte-Carlo
simulations of the Boltzmann equation for the occupations of the k-states in
the individual subbands
\cite{TortoraPHB1999,IottiAPL2001,WaldmullerIEEEJQuant2006,JirauschekJAP2007},
density matrix calculations
\cite{CallebautJAP2005,KumarPRB2009,GordonPRB2009,DupontPRB2010,TerazziNJP2010},
which have been also done k-resolved
\cite{IottiPRL2001,WaldmullerIEEEJQuant2006,WeberPRB2009},  as well as
nonequilibrium Green's functions (NEGF)
\cite{LeePRB2002,SchmielauAPL2009,KubisPRB2009,HaldasIEEEJQuant2011,GrangePRB2015}.
While the published results from either scheme typically agree well with
experimental data, it is not clear how the choices of parameters (in
particular interface roughness (IFR) distributions and band offsets), specific
approximations (such as screening models or various model-specific assumptions as
subband temperatures), or model complexity affects the results.

Therefore we performed simulations with our NEGF scheme
\cite{WackerIEEEJSelTopQuant2013} for a wide range of different published
THz QCLs using precisely the same parameters and model approach, and
document \textit{all} results in this paper.  This allows to monitor the
quality of our simulation scheme. Furthermore, the comparison between devices
from different groups can reflect systematic trends. Here it is well-known,
that the reproduction of devices from different groups frequently provided
different results, where the origin is far from understood, see, e.g.,
Ref.~\onlinecite{ChanAPL2013}. 
It is also known that samples grown at the same lab but under different
growth campaigns can differ, although methods now exist to guarantee 
run-to-run reproducibility\cite{LiOE2015}. 
The identification of trends in published samples from different
labs may shine light into discrepancies of the growth procedures,
and to pose the right questions for the community to take steps towards 
inter-lab reproducibility of THz QCL devices. 

\section{Model and estimates} \label{SecModel}
One of the most important parameter in heterostructure modeling is the
conduction band offset (CBO), a function of the bandgap of the alloys in
question and the valence band offset (VBO).  In this work we limit ourselves
to ${\rm Al}_x{\rm Ga}_{1-x}{\rm As}/{\rm Ga}{\rm As}$ systems in the direct
bandgap regime.  Consulting the standard literature, Vurgaftman \textit{et
  al.}\cite{VurgaftmanJAP2001}  provides the relation ${\rm CBO} = 0.97 x$ eV
(in the vicinity of 15\% Al content; the full expression is  a cubic
polynomial in $x$) for ${\rm Al}_x{\rm Ga}_{1-x}{\rm As}$ barriers. This is
in turn based on a VBO of $0.53 x$ eV, assumed temperature independent,  as
well as low temperature results for the band gaps of  both GaAs and AlAs. This
relation for the conduction band offset is however seldom used in the QCL
community \cite{SirtoriIEEEJQuant2001,KumarDiss2007,ChanAPL2013}, where
instead a lower offset  is often preferred.  This might be more reasonable for
the design of structures aimed at high temperature operation, where we expect
the band gap to decrease.  More recent experiments by Yi \textit{et
  al.}\cite{YiPRB2010} and  Lao \& Perere\cite{LaoPRB2012} have found low
temperature VBOs of $\sim 0.570 x$ eV, which  would distribute more offset to
the valence band side,  effectively lowering the CBO. This justifies the use
of a lower value compared to Vurgaftman \textit{et al.}.  In total, Yi
\textit{et al.} found a CBO of $0.831 x$ eV for $x \le 0.42$, at 4.2 K.  As the
focus of this study is the performance of THz QCL at low temperatures, we will
use this result for the CBO in the remainder of this work.  Furthermore we use
the effective mass of the conduction band edge  $m_{eff}=0.067 + 0.083x$, as
given in Vurgaftman \textit{et al.}\cite{VurgaftmanJAP2001} and used by most groups.  Together with
standard material parameters for bulk GaAs, this defines the  heterostructure
apart from the doping density and layer sequence. The basis  states are then
calculated in an effective two band
model\cite{LindskogSPIE2013,FranckieOE2015} using a Kane energy of $E_P=22.7$
eV\cite{KaneJPhysChemSolids1957}. 

In our transport model we calculate the scattering self-energies in the
self-consistent Born approximation. For the structures of interest, the
elastic processes are dominated by impurity and IFR scattering. 
In addition we include alloy scattering to the elastic self-energies.
For modeling IFR we use an exponential correlation function in this
work. Gaussian correlation functions have been shown to yield similar results
\cite{FranckieOE2015}, and the values used here, $0.2/10.0$ nm for rms
height/correlation length can best be compared to $0.2/7.0$ nm for a Gaussian
correlation function. Here the use of an effective two band model decreases
the impact of the IFR scattering \cite{FranckieOE2015}.

In their complete theoretical formulation, the self energies of the NEGF
models are functions of both momentum and energy, but in our implementation
they are effectively treated as only energy dependent. This is done by
evaluating the scattering matrix elements at a set of representative momentum
transfers\cite{WackerIEEEJSelTopQuant2013}, and reduces the  computational
effort to an accessible level. 
Throughout this work we use the typical momentum transfers $k_0$ equivalent to 
$E_{k}^0=6.3$ meV for intra-subband scattering at 77 K lattice temperature.
For intersubband scattering, we also adjust for the difference in subband energy.
One problem with this procedure is that it introduces a logarithmic
divergence in the real parts of the self-energies for increasing $k$-ranges
covered. The choice of typical momentum transfers and our solution 
to the problem of the divergence is discussed in detail in Appendix~\ref{AppDivFix}. 

For the inelastic self-energies we include both acoustic and longitudinal
optical (LO) phonon scattering. Electron-electron scattering is also implemented via a
rudimentary form of the GW approximation\cite{WingeJPhysConfSer2016}, where
the screening function is replaced by its plasmon pole. This allows us to go
beyond the meanfield approximation and to estimate in which type of structures
this mechanism is of importance. The modeling is based on energy exchange of
the conduction band electrons with a plasmon bath, with a temperature equal to
an effective electron temperature, in order not to artificially cool the
electron gas. In Appendix~\ref{AppElectronTemp} we discuss how this
temperature can be chosen by balancing the electric power dissipated in the
structure and the cooling rate of the LO phonons.

\section{Samples studied} 
In the following, we present a short overview of each class of designs
included in this work, and introduce and label each device. They will be
referenced in the text by a shorthand notation containing first author name,
journal and year, e.g. DupontJAP2012 for the device considered in
Ref.~\onlinecite{DupontJAP2012}.

\subsection{2-well designs}
The 2-well design is the simplest possible realization of a QCL, which next to
the upper and lower laser level employs one further level, serving both for
extraction from the lower laser level (by LO phonon scattering) and  injection
by resonant tunneling (RT) into the upper laser level of the next period.
Here we study the first realized structure KumarAPL2009B\cite{KumarAPL2009B}
as well as the broadband laser ScalariOE2010\cite{ScalariOE2010}. Both lasers
were processed with metal-metal (MM) waveguides and the latter showed high power
output when parts of the upper metal contact was removed. 

\subsection{3-well designs}
The 3-well structures apply an additional well for extraction and thus employ
a RT LO phonon depletion mechanism for emptying the lower laser
state\cite{WilliamsAPL2003}. This type of design has achieved high operation
temperatures with a record temperature achieved in
2012\cite{FathololoumiOE2012}.  Variants have been realized in all labs
included in this study, except ETH. This fact and the  simple layer sequence
strategy makes it a very interesting type of design when we want to  compare
samples of different origin.  The lasers studied here are
FathololoumiOE2012\cite{FathololoumiOE2012}, sample V812 of
FathololoumiJAP2013\cite{ FathololoumiJAP2013},
KumarAPL2009A\cite{,KumarAPL2009A}, DeutschAPL2013\cite{DeutschAPL2013} and
SalihJAP2013\cite{SalihJAP2013}.  The designs are similar in principle, with
small changes in doping densities and oscillator strengths.  Despite this
similarity, the reported output powers differ  drastically depending on both
origin and growth campaign as seen in TAB.~\ref{TabSummary}.  This is most
probably due to the procedure of removing parts of the top contact layer of
KumarAPL2009A and FathololoumiOE2012, a procedure described in
Ref.~\onlinecite{BelkinOE2008},  which reduces waveguide losses for these
samples.  

\subsection{Hybrid designs}
The class of structures that we will denote as \textit{hybrid} borrows ideas
both from the bound to continuum structures, which were the first to lase in
this spectral range\cite{KohlerNature2002}, and the 3-well concept. In these
structures a fourth well is inserted, and extraction from the lower laser
state to the extractor state occurs through a combination of scattering and
RT.  Here we study two subclasses of this design type. In the
designs BurghoffAPL2011\cite{BurghoffAPL2011}, MartlOE2011\cite{MartlOE2011}
and BenzAPL2007\cite{BenzAPL2007} a two well injector is used, similar to the
pioneering design of Ref.~\onlinecite{WilliamsAPL2003}, whereas the other
samples  use RT injection directly from the ground state of
the phonon well into the  upper laser state as proposed in
Ref.~\onlinecite{AmantiNJP2009}.  Showing good scaling properties, the latter
designs are suitable for high power operation\cite{LiElectronLett2014}, and
the robustness in layer sequence can be utilized for broadband multi-stack
devices\cite{TurcinkovaAPL2011}.  Here, these designs are labeled as
LiEL2014\cite{LiElectronLett2014}, AmantiNJP2009\cite{AmantiNJP2009} and
TurcinkovaAPL2011\cite{TurcinkovaAPL2011}, respectively.  As the layer
sequence of stack A in Ref.~\onlinecite{TurcinkovaAPL2011} is identical  to
sample EV1157 of Ref.~\onlinecite{AmantiNJP2009}, we compare here to results
of the high doped version, labeled N907, when we refer to AmantiNJP2009. 

\subsection{Indirectly pumped designs}
We have also studied the indirectly pumped, also known as scattering assisted
injection\cite{KumarNatPhys2011} and phonon-photon-phonon (3P)
designs\cite{DupontJAP2012}. These designs use LO phonon scattering to
populate  the upper laser state, which is fundamentally different to the RT
injection but requires a larger bias to operate. Here we present results for
DupontJAP2012\cite{DupontJAP2012}, RazavipourJAP2013\cite{RazavipourJAP2013}
and KhanalOE2015\cite{KhanalOE2015}. 

\begin{table*}
   \centering
   \begin{tabular*}{\textwidth}{c @{\extracolsep{\fill}} lccccccrrcccc}
   Type &  Ref. & $J_\textrm{exp}^{\rm thr}$ & $J^{\textrm{thr}}_\textrm{sim} $ & $J_\textrm{exp}^{\rm peak}$ & $J^\textrm{dc}_\textrm{sim}$ & $J_\textrm{sim} ^{\rm lase}$ & $J_\textrm{sim}^{\textrm{GW dc}}$ 
   & $I_{\rm exp}$  & $I_{\rm sim}$ & $\nu_{\rm exp}$ & $n_{2D}$ & $g_{th}$ & Origin  \\
    \hline \hline
    \multirow{2}{*}{2-well} & KumarAPL2009B \cite{KumarAPL2009B} & 415 & 510*
    & 950 & 1040 & 1700 & 1100 & -- & 2200 & 4.5 & 2.2 & 20 & Sandia \\
    & ScalariOE2010     \cite{ScalariOE2010}                 & 470 & 440
    &  800 &  520 &  780* &  530 & 290 & 790  & 3.2 & 1.5 & 15 & ETH \\
    \hline
    \multirow{5}{*}{3-well}  & KumarAPL2009A \cite{KumarAPL2009A} & 440 & 500
    &  850 &  610 & 1860 &  920 & 580  & 1460 & 3.9 & 3.0 & 20 & Sandia \\
    & SalihJAP2013 \cite{SalihJAP2013}                         & 1300 & 1050*
    & 1400 & 1300 & 1300 & 1500 & 0.5 & 22 & 3.2 & 2.75 & 100 &  Leeds \\
    & FathololoumiOE2012         \cite{FathololoumiOE2012}          & 1000 & 1200
    & 1600 & 1350 & 2120 & 1800 & 275 & 1260 & 2.7 & 3.0 & 15 &  Ottawa \\
    & FathololoumiJAP2013  \cite{FathololoumiJAP2013}        & 660 & 715
    & 1000 &  900 & 1490 & 1270 & 60 & 1240 & 3.3 & 3.0 & 15 & Ottawa \\
    & DeutschAPL2013 \cite{DeutschAPL2013}                       & 900 & 690
    & 1400 &  850 & 1230 & 1075 & 70  & 940 & 3.8 & 3.3 & 20 & Vienna \\
    \hline 
    \multirow{6}{*}{Hybrid} & BurghoffAPL2011 \cite{BurghoffAPL2011} & 360 & 670 
    & 420 & 760 & 970 & 1030  & --  & 790  & 2.2 & 3.0 & 10 & Sandia \\  
    & LiEL2014\cite{LiElectronLett2014}                             & 520 & 470
    &  700 &  500 & 1100 & 1240 & 250 & 340 & 3.4 & 5.2 & 40 & Leeds \\ 
    & AmantiNJP2009     \cite{AmantiNJP2009}            & 450 & 680
    & 810 &  740 & 1045 & 1250  & 20  & 470  & 3.0  & 11 & 15 & ETH \\ 
    & TurcinkovaAPL2011 \cite{TurcinkovaAPL2011}                   & 300 & 215  
    &  430 &  225 &  500 &  670 & 50 & 220 & 3.0 & 3.7 & 45 & ETH \\
    & BenzAPL2007   \cite{BenzAPL2007}                        & 510 & 450
    &  820 &  590 &  770 &  540 & --  & 385 & 2.8 & 1.9 & 15 & Vienna \\
    & MartlOE2011 \cite{MartlOE2011}                          & 165 & 125*
    &  215 &  145 &  170* &  170 & --  & 130 & 2.1 & 0.6 & 10  & Vienna \\
    \hline
    \multirow{3}{*}{Indirect} & KhanalOE2015 \cite{KhanalOE2015} & 800 & 500*
    & 1350 &  850 &  900 &  740 & 130  & 460 & 2.1 & 3.17 & 10 & Sandia \\
    & DupontJAP2012\cite{DupontJAP2012}                        & 1250 & -- 
    & 1600 & 1800 & -- & 2000 & 10 & 0 & 3.0 & 3.25 & 15 & Ottawa \\ 
    & RazavipourJAP2013 \cite{RazavipourJAP2013}               & 850 & 1100*
    & 1300 & 1490 & 1360 & 1600 & 85   & 1300  & 2.4 & 3.45 & 10 & Ottawa \\ 
    \hline \hline
   \end{tabular*}
   \caption{Collected simulation results together with the most relevant 
experimental measurements, grouped with respect to design and origin. 
Current densities denoted by $J$ are given in A/cm$^2$, measured 
frequencies $\nu_\mathrm{exp}$ in THz, sheet doping densities $n_{\rm 2D}$
in $10^{10}$~cm$^{-2}$ and threshold gain $g_{th}$ in 1/cm. 
Calculated currents showing a particular large sensitivity 
to the value $g_{th}$ are marked by asterisks. The simulated lasing intensity within the active region 
$I_{\rm sim}$ has units [$\mu$W/$\mu m^2$]. The experimental counterpart $I_{\rm exp}$  is based on
an estimated collection efficiency of 30\% of the reported power 
at low temperature pulsed operation. A graphical display of the results is found in 
FIG.~\ref{FigComp}. }
\label{TabSummary}
\end{table*}

\section{Procedure}

For each device, we performed simulations based on the nominal sample
parameters as listed in each publication. The input parameters are the 
reported sheet doping densities and layer sequences,
whereas the roughness and all other scattering parameters were kept the same 
(For comparison of the intensity inside the waveguide, also the sample facet area 
was read out). 
We use 77 K as lattice temperature in all simulations and 
compare to experimental data taken in pulsed mode operation, 
at heat-sink temperatures close to 77 K when possible, 
or at lower temperatures otherwise. Usually, the
experimental data do not show any significant variations in this temperature
range.

The main results are collected in TAB.~\ref{TabSummary} together with some basic data of the devices, such as the sheet
doping density per period $n_{2D}$ and the reported lasing frequency. Most
importantly, we provide the experimental peak current density $J_\textrm{exp}^\textrm{peak}$
under laser operation for low temperatures, which is the key quantity for
comparison, and also the threshold current $J_\textrm{exp}^\textrm{thr}$. 
For comparison, we provide the $J_\textrm{sim}^\textrm{thr}$, where the simulated gain reached the threshold value,
and the peak current $J_\textrm{sim}^\textrm{dc}$ of our simulations neglecting the ac field of lasing operation.

A key parameter for the simulation is the
threshold gain $g_\textrm{th}$ in the heterostructure required to overcome
waveguide and mirror losses.  For the MM waveguides operating around 2 THz  we
use $g_\textrm{th}=10/{\rm cm}$. This roughly corresponds to the electric field
losses (which are half the intensity loss used here) of 4.3/cm addressed in 
Ref.~\onlinecite{MartlOE2011}. For  frequencies above  2.7 THz we use
$g_\textrm{th}=15/{\rm cm}$ and above 3.8 THz we use $g_\textrm{th}=20/{\rm cm}$ 
for the MM waveguides, taking into account the higher
attenuation in the metals. The device TurcinkovaAPL2011 is a multi-stack
design, and as each stack has a significantly smaller mode confinement factor
compared to standard devices, we use the enhanced value of
$g_\textrm{th}=45/{\rm cm}$.  For the semi-insulating surface-plasmon (SI-SP) 
waveguide in LiEL2014, we use
$g_\textrm{th}=40/{\rm cm}$ guided by results given in
Ref.~\onlinecite{KohenJAP2005}. For the  second  SI-SP waveguide sample
SalihJAP2013, we use $g_\textrm{th}=100/{\rm cm}$ based on the calculated 
waveguide losses in the publication for a 6 $\mu$m wide active region. 
Note that the free carrier
loss in the active region itself is taken into account in the NEGF simulations
of  the gain spectra.

The response of the active region under operation 
is simulated using a classical ac field with increasing strength 
until gain is saturated to the threshold value. This provides an increased peak current $J_\textrm{sim}^\textrm{lase}$.
The ac field strength is then related to the intensity inside the 
active region $I_{\rm  sim}$ via the Poynting vector. The corresponding 
experimental value $I_{\rm  exp}$ is obtained from the measured power output  at low temperature and pulsed
mode operation in the following way: The power is divided by a typical collection efficiency of 
30\% \cite{AjiliAPL2004,VitielloAPL2007,BelkinOE2008} and the transmittivity leaving the waveguide,
multiplied with the confinement factor of the waveguide, and finally divided by the cross section of the active region.
For the transmittivity we use 0.25, 0.20 and 0.15 for the MM waveguides 
with $g_{th}$ of 20, 15 and 10 per cm, respectively, guided by the results
of Ref.~\onlinecite{KohenJAP2005}. 
For the SI-SP waveguides we use a transmittivity of 0.68 calculated 
from the Fresnel equations. We use unity confinement 
factors for the MM waveguides, 0.4 for the waveguide of LiEL2014\cite{confinementfactor} and 
0.2 for the waveguide in SalihJAP2013\cite{SalihJAP2013}.
In this context we note, that the gain saturation, defining the simulated intensity $I_{\rm  sim}$ contains 
backwards traveling waves, which do not contribute to the experimental output
as discussed in Ref.~\onlinecite{WingeOE2014}. Thus, $I_{\rm  sim}$ is expected to overestimate the output by
a factor up to two for low transmittivity\cite{backtravel}.  These considerations show, that the values of $I_{\rm  exp}$ can only be seen as rough estimates, where a large part of the uncertainty is due to the collection efficiency. 

\begin{figure}
\centering
\includegraphics[width=1.0\columnwidth]{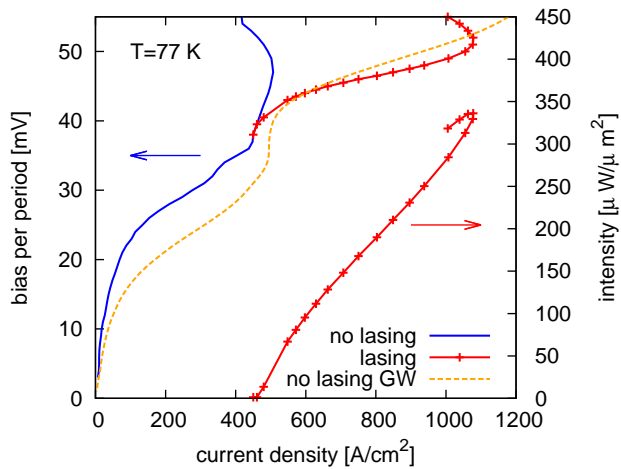}
\caption{Light-current-voltage (LIV) characteristics of the 1~W THz QCL presented in Ref.~\onlinecite{LiElectronLett2014} (LiEL2014). 
The solid blue line indicate the current without lasing, while the crossed red lines show current and intensity under operation. 
The dashed orange line refers to the simulations taking into account
electron-electron scattering in the plasmon-pole approximation.}
\label{FigLiLIV}
\end{figure}

\begin{figure}
\centering
\includegraphics[width=0.9\columnwidth]{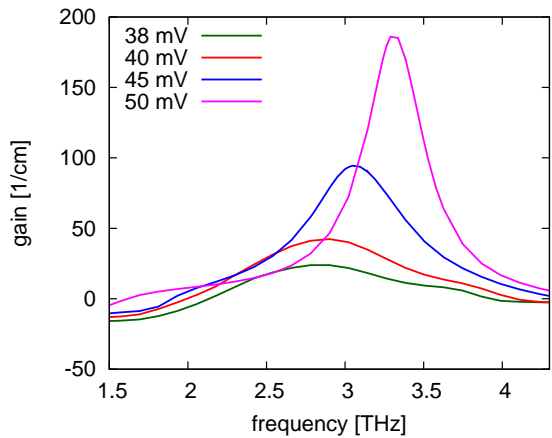}
\caption{Gain simulations at a number of different bias points corresponding to the LIV in FIG.~\ref{FigLiLIV}. 
A Stark shift can be observed with increasing bias.
Experimentally lasing was reported around 3.4 THz
\cite{LiElectronLett2014}.}
\label{FigLiGain}
\end{figure}

As an example for our simulations, we show more detailed results for
the hybrid design of LiEL2014\cite{LiElectronLett2014} in FIG.~\ref{FigLiLIV}.
The dc calculations without the ac field  (full blue line) exhibit a peak current density
$J_\textrm{sim}^\textrm{dc}=500 \, \mathrm{A/cm}^2$, and this value is presented
in the sixth column of TAB.~\ref{TabSummary}.  Then we consider the spectral
gain for different biases as shown in FIG.~\ref{FigLiGain} in order to
determine the threshold current $J^\mathrm{thr}_\textrm{sim}$, 
where the material gain surpasses the
threshold gain $g_\mathrm{th}$ (if this operation point is in a region of
negative differential conductivity, we assume that the threshold occurs in a
domain state and use the current density of the preceding peak).  
Above the threshold current, we perform
simulations in the presence of the ac field in order to study the operating
device. Here we increase the ac field until gain saturation reduces the gain to $g_\mathrm{th}$.
This provides the light-current-voltage (LIV) characteristics shown in
FIG.~\ref{FigLiLIV} by a red line with crosses. The maximum current achieved
is denoted by $J_\textrm{sim}^\textrm{lase}$ (here $1100 \, \mathrm{A/cm}^2$) and
this value is presented in the seventh column of TAB.~\ref{TabSummary}. In the
same way we obtain the intensity $I_\textrm{sim}$ displayed in the tenth column
of TAB.~\ref{TabSummary}. In order to judge the relevance of
electron-electron scattering, we performed simulations with our
plasmon-pole approximation for the off state, as shown by the orange dashed line
in FIG.~\ref{FigLiLIV}. The corresponding peak current is denoted as
$J^\textrm{GW dc}_{\rm sim}$.


\begin{figure}
\centering
\includegraphics[width=0.9\columnwidth]{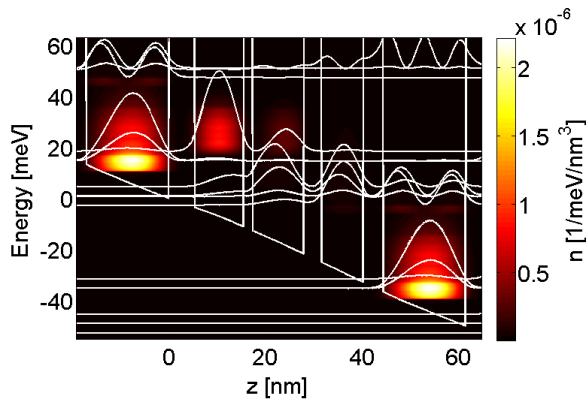}
\caption{Energy resolved densities calculated from the lesser Green's function showing how the populations are distributed at the design bias of 50 mV per period, without any laser field included. The inversion between the upper and lower laser state is clearly visualized.}
\label{FigLiDens}
\end{figure}

Now we discuss the specific results for the sample
LiEL2014\cite{LiElectronLett2014} in more detail: Around the bias matching
the optical  phonon energy of $\sim36$ meV the parasitic injection channel is
enhanced and seen to give a small feature in the current. After this, the
upper laser state is instead favored by the tunneling transition, and
inversion is  building up, allowing for laser action to start provided the
losses are low enough.  With no lasing in the cavity, the peak current
saturates at 500 A/cm$^2$ before the design bias of 50 mV is reached. The
situation is displayed in FIG.~\ref{FigLiDens} where the electron  densities
are resolved in energy and growth direction. Inversion is clearly visible, but
without laser field the electrons stay relatively long in the injector and
upper laser state.  When the structure is modeled with a laser field, current
is enhanced and the negative differential resistance (NDR) feature is shifted to 52 mV, allowing the structure
to reach its intended configuration.  From the GW results in
FIG.~\ref{FigLiLIV} we see also that the inelastic scattering from
electron-electron interactions can enhance the charge transfer through the
structure.

Assuming losses of 40/cm, the threshold current is 470 A/cm$^2$, which is in
reasonable agreement with  the experimental value at ~400 A/cm$^2$. The
calculated dynamical range however, can be seen to exceed 600 A/cm$^2$ which
is a factor of two larger than what is seen experimentally. Simultaneously,
the calculated intensity $I_\mathrm{sim}$ is slightly above the experimental value
$I_\mathrm{exp}$. 
Here we also note, that we assume a homogeneous field ac and dc
field in the heterostructure region of the sample. While this is
well-justified for the dc-field, the actual ac-field-distribution depends on
the waveguide. In particular for SI-SP waveguides, the ac-field is not
homogeneous in the growth direction, which is disregarded by our scheme. Thus we
are not surprised, that our approach overestimates both the simulated peak
current $J_\textrm{sim} ^{\rm lase}$ and intensity $I_{\rm sim}$.

\section{Results}
Collected results from all simulations are shown in TAB.~\ref{TabSummary}
together with relevant experimental quantities. In the following, we focus on
the  maximum current under operation and the threshold current, which are most
easily extractable for the experimental data. (Only two samples, SalihJAP2013
and MartlOE2011, did not show a clear NDR feature after maximum power in their
respective reference,  which adds some uncertainty in their respective
$J_\textrm{exp}^\textrm{peak}$.)  As the assumption of homogeneous ac-fields
tends to overestimate the lasing intensity, we would expect this experimental
peak current to be between $J_\textrm{sim}^\textrm{dc}$ and
$J_\textrm{sim}^\textrm{lase}$. If  $J_\textrm{sim}^\textrm{GW dc}$ is much
larger than $J_\textrm{sim}^\textrm{dc}$, electron-electron scattering appears
to be more important, so that the other calculated currents may be too low.

In most of our simulations, the peak current is a slowly varying function of
the losses once high intensity is reached, which reduces the impact of an
erroneous estimate of $g_{th}$. On the other hand, the threshold current can
be  very sensitive to this parameter. For samples where we find that the
currents are very sensitive  to the value of $g_{th}$, we mark the relevant
quantity with an asterisk in TAB.~\ref{TabSummary}.

\begin{figure}
\centering
\includegraphics[width=0.95\columnwidth]{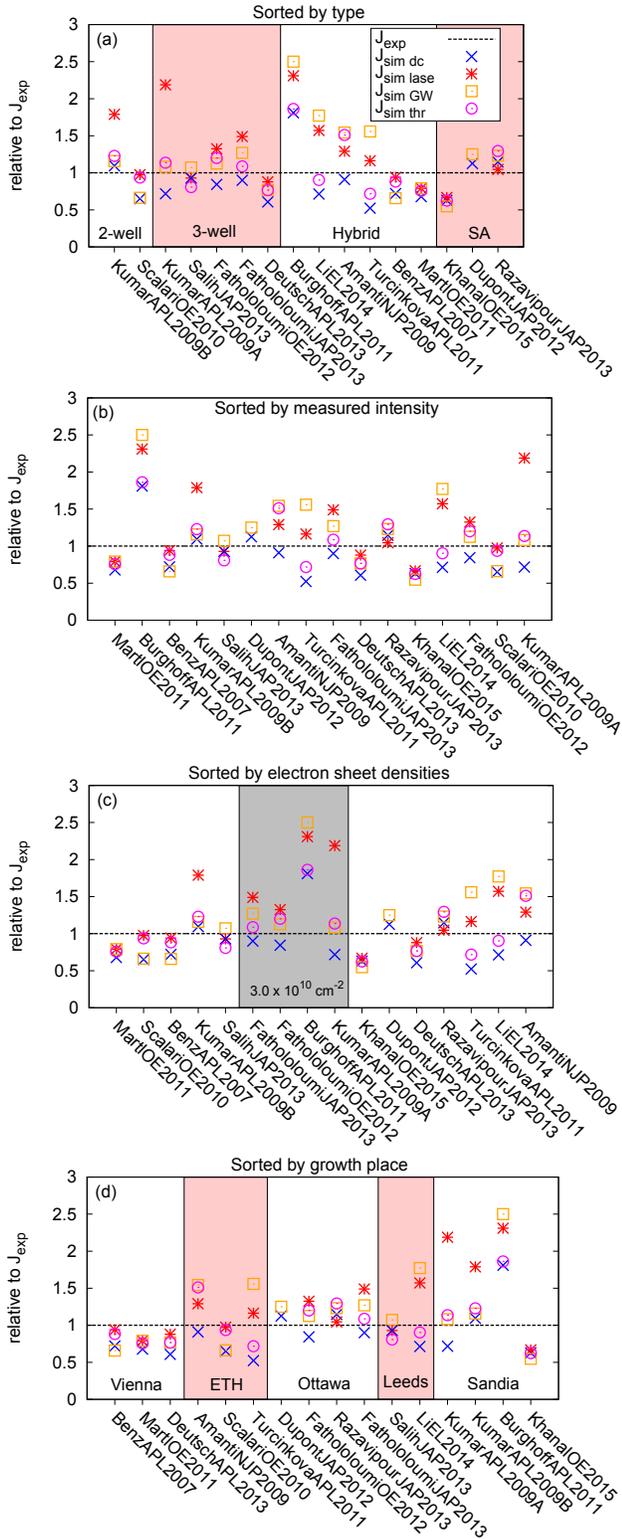}
\caption{Simulated currents divided by experimental currents. 
The panels order the samples according to (a) design class as in TAB.~\ref{TabSummary}, (b) from low (left) to high (right) experimental lasing intensity, (c) from low (left) to high (right) sheet doping density, and in (d), samples from the same lab have been grouped together. }
\label{FigComp}
\end{figure}

The ratios $J_\textrm{sim}^\textrm{dc}/J_\textrm{exp}^\textrm{peak}$,
$J_\textrm{sim}^\textrm{lase}/J_\textrm{exp}^\textrm{peak}$,
$J_\textrm{sim}^\textrm{GW dc}/J_\textrm{exp}^\textrm{peak}$, and
$J_\textrm{sim}^\textrm{thr}/J_\textrm{exp}^\textrm{thr}$ are displayed in
FIG.~\ref{FigComp}(a) for an easy identification of the overall quality of
the simulations. As in TAB.~\ref{TabSummary}, the samples are ordered
according to the design class. We find, that the model provides good results
for many samples, but within each design class there are devices, where the
experimental and calculated currents disagree significantly.  Furthermore, we
find, that $J_\textrm{sim}^\textrm{GW dc}$ does not differ much from
$J_\textrm{sim}^\textrm{dc}$ except for several hybrid designs, such as
LiEL2014 discussed above. 

In order to study, why the simulations appear to describe some devices better
than others, we now order the results in different ways. First, we order the
devices according to the measured intensity, which could reveal problems of
our model to describe the devices under operation. However,
FIG.~\ref{FigComp} (b) shows no clear trend. Secondly, we order the devices
according to the doping intensity, which is relevant for impurity scattering
and electron-electron scattering. Again, FIG.~\ref{FigComp} (c) does not
provide any trend for the reliability of our model (samples with sheet doping density
of $3 \cdot 10^{10} / {\rm cm}^{2}$ are shaded as a guidance). However, we note, that
for high doping density  $J_\textrm{sim}^\textrm{GW dc}$ becomes much larger
than  $J_\textrm{sim}^\textrm{dc}$ indicating the relevance of
electron-electron scattering. Thirdly, we  sort the devices with respect to
growth place in FIG.~\ref{FigComp}(d). Here we find a clear trend, where 
the simulations with the parameters mentioned above provide too low currents for the devices grown at the Technical
University of Vienna and too high currents for devices grown at Sandia (except
for KhanalOE2015,  which we discuss below).  For samples grown at the NRC in
Ottawa and in Leeds, our model provides good agreement with the experimental
peak current and threshold current.  For samples grown at ETH, the results are
slightly more scattered. In particular the calculated
$J_\textrm{sim}^\textrm{thr}$ for AmantiNJP2009 is too large.

The rightmost device in FIG.~\ref{FigComp}(d) KhanalOE2015, an indirect design, does not fit the
picture. While the indirect designs have barriers with  $x=0.25$, this device
has a lower barrier height ($x=0.15$), which may provide substantial leakage
into the  continuum\cite{ChanAPL2013} for the high electric field required for
indirect designs. This is not taken into account in the NEGF model and could
explain, why the model provides a smaller peak current. To explore this
hypothesis, we have studied the parasitic resonances where they cause an
experimental NDR feature, visible as a distinguished pre-peak in the LIV.  This
is true for all studied samples of the indirect class,  and also for
KumarAPL2009A, KumarAPL2009B, FathololoumiOE2012, DeutschAPL2013 and
FathololoumiJAP2013.  For the samples from Sandia, including KhanalOE2015, we overestimate the pre-peak
currents by more than 25\% in all cases, while we have agreement within 20\%
for the other samples. Thus, the currents at the parasitic resonances agree
with the main trend of FIG.~\ref{FigComp}(d), that we overestimate the
currents of the Sandia devices. 

In addition we compare simulated and measured intensity for each sample.  The
procedure is described in Sec.~\ref{SecModel}, and experimental values, when
available, are listed in TAB.~\ref{TabSummary}. 
For all structures studied, we
overestimate the waveguide intensity. For the sample LiEL2014, utilizing a 
SI-SP waveguide, we get rather good agreement between our
calculations and the experimental data. Assuming, that backwards traveling
waves correspond to half the calculated intensity in the MM
waveguides of ScalariOE2010, KumarAPL2009A, KhanalOE2015, the
discrepancy is reduced to less than 50\%. The other samples show even lower measured
intensity, which we cannot explain here. Possible causes could be the
overestimation of the simulated intensity due to simplified model assumptions
such as the neglect of heated phonons \cite{VitielloAPL2012,ShiJAP2014},
uncertainties in parameters such as threshold gain and
transmittivity, an experimental collection efficiency far below $\sim30\%$,
or non-uniformity of the laser field in the cavity.

\section{Sensitivity of growth parameters}
The trends observed for samples from different groups indicates, that samples grown from same the design at different places 
are not identical. Likely reasons are deviations in doping, the Al content $x$, as well as different IFR.
In our simulations we normally observe
a linear increase in current with doping, although the same trivial dependence does not hold for gain 
\cite{WackerAPL2010,GrangePRB2015}. 
As the dependence on IFR and Al content are not equally well understood, we have conducted several numerical 
experiments to estimate the impact of fluctuations in these parameters.

\begin{figure}
\centering
\includegraphics[width=\columnwidth]{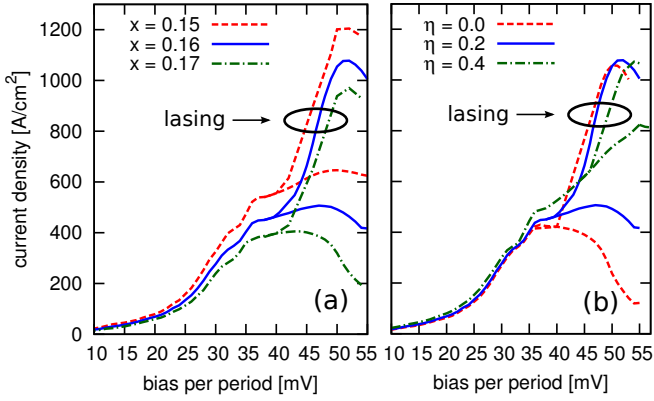}
\caption{Sensitivity of currents to varying simulation parameters for
    the sample LiEL2014\cite{LiElectronLett2014}. 
     In (a) the Al content $x$ is changed giving lower and higher barriers for
    $x=0.15$ and $x=0.17$ respectively,  compared to the nominal one
    with $x=0.16$. In (b) the  IFR height $\eta$ is changed 
    within two extreme values.}
\label{FigLiXIFR}
\end{figure}

By increasing the  barrier height we expect a decrease in current as the tunneling amplitudes 
decrease. This is indeed what we find, and we show this in FIG.~\ref{FigLiXIFR}(a) where
simulations with altered Al content compared to the nominal simulations in FIG.~\ref{FigLiLIV}
are shown. The linear dependence of both dc and lasing current show few surprises; 
here decreasing barrier heights of
1\% Al gives an increase in current of about 10\%. 
In contrast, the impact of changing the rms roughness height $\eta$ is more remarkable as shown in FIG.~\ref{FigLiXIFR}(b). 
Compared to the nominal calculations with $\eta=0.2$ 
we get 50\% more dc current when $\eta$ is doubled. On the other hand,
the current under irradiation is almost unchanged. 
This conservation of current can be understood as two competing mechanisms, where either 
stimulated photon emission or elastic scattering is depleting the upper laser state. 
The cost of stronger scattering is a smaller dynamic range
which will impede higher temperature operation. 
This analysis suggests two bottlenecks in the transport,
the lifetime of the upper laser state and the tunneling rate over the injection
barrier.

\begin{figure}
\centering
\includegraphics[width=\columnwidth]{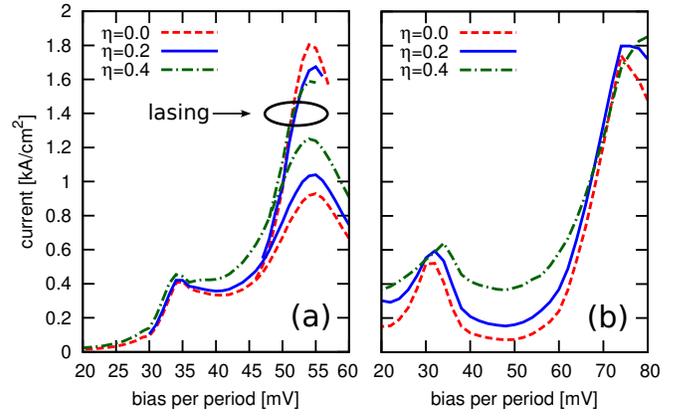}
\caption{Simulated current for different IFR heights $\eta$ for the samples 
KumarAPL2009B (a) and DupontJAP2012 (b). }
\label{FigDupKum}
\end{figure}

In order to verify this observation, we performed corresponding simulations with modified IFR for two further structures with different designs
as shown in FIG.~\ref{FigDupKum}. 
For KumarAPL2009B, FIG.~\ref{FigDupKum}(a) confirms the trend seen for LiEL2014 in FIG.~\ref{FigLiXIFR}(b),
that the current under lasing balances the drastic changes seen in the dc current when the roughness 
parameters are changed. Here the peak current under lasing also shows some sensitivity to the roughness
parameters. We believe this to be the coherent part of the injection tunneling current 
decreasing due to increased scattering. This can explain the loss in total current (incoherent and coherent) under lasing for increasing roughness, as the upper laser state is no longer as efficiently populated. 
In the case 
of DupontJAP2012 we see a slight increase at lower biases with increasing scattering, as
shown in FIG.~\ref{FigDupKum}(b), however at design bias, 
the current is not significantly effected, and the current at both peaks agrees well with the experimental data.
As this laser is depopulated via RT
and subsequent resonant phonon scattering\cite{DupontJAP2012}, 
this current bottleneck seems not be be very 
sensitive to additional elastic scattering. 
These simulations provided gain below 10/cm for all roughness heights at the
current peak. However, including our rudimentary electron-electron scattering
or raising temperature provides gain slightly above 15/cm, in accordance with
the observed weak lasing.

\section{Conclusions}
An extensive study was made including 16 samples among published work
on terahertz QCLs from the last ten years, using our NEGF simulation
scheme. Using identical simulation parameters, we find that the simulated
current does not agree with experimental results for all samples. However,
we observe a clear trend that these deviations are similar for samples
form a given laboratory. This shows that samples from different
laboratories are not fully comparable. We show that interface roughness
alone cannot account for these deviations in the simulated current under
lasing compared to experimental data. Assuming different calibrations of
doping density or Al content in different laboratories could explain these
trends. However more intricate issues, such as different barrier profiles,
cannot be ruled out.

\begin{acknowledgements}
We thank K. Unterrainer, G. Strasser, and J. Faist for helpful discussions and 
the Swedish Research Council for financial support.
\end{acknowledgements}

\appendix
\section{Asymptotic behavior of the self-energies} \label{AppDivFix}
In the formalism presented in Ref.~\onlinecite{WackerIEEEJSelTopQuant2013} the elastic
self-energies are expressed as 
\begin{align} \nonumber
&\Sigma_{\alpha\alpha'}^{\rm </r}(E,E_k)= 
\sum_{\beta\beta'}
\int_0^{\infty}\d E_{k'} G_{\beta\beta'}^{\rm </r}(E,E_{k'}) \times \\ \label{EqSigma}
&\underbrace{\frac{\rho_0A}{4\pi}
\int_0^{2\pi}
\d \varphi 
\langle V_{\alpha\beta}(E_k,E_{k'},\varphi)
V_{\beta'\alpha'}(E_k,E_{k'},\varphi)
\rangle_{\rm imp}}_{X^{\rm elast}_{\alpha\alpha',\beta\beta'}(E_k,E_{k'})}  
\end{align}
where \textless/r denotes the lesser and retarded objects, respectively, $\rho_0$ is
the background density $em_{eff}m_e/(\pi\hbar^2)$, $A$ is the lateral area and
$V_{\alpha\beta}$ are the scattering matrix elements for each process considered, impurity 
averaged over all configurations of scattering potentials.
The self-energies, being functions of both $E$ and $E_k$, are integrals in $E_{k'}$
 of the Green's function $G(E,E_{k'})$ and the angle averaged matrix elements $X_{\alpha\alpha'\beta\beta'}^{\rm elast}(E_k,E_{k'})$. 

Using the non-interacting Green's function $G_{\beta\beta}(E,E_{k'})\sim(E-E_\beta-E_{k'}+\imai\Gamma)^{-1}$ 
with a phenomenological broadening $\Gamma$, as an approximation, the most interesting parts in the self-energies are found around 
$E=E_\beta+E_{k'}$ and at $E=E_\alpha+E_k$, where the self-energies peak. 
This motivates the use of the typical energies $E_{k}^0,E_{k'}^0$, effectively moving the 
scattering matrix elements out of the integral over $E_{k'}$ and making the 
self-energies functions of $E$ only.
To fix $E_{k}^0$, we calculate the intra- and inter-subband scattering rates
using thermalized Boltzmann-like subbands at different electron temperatures.
This is done for a set of representative low doped heterostructures, and 
the $E_k^0$ giving the best agreement for the self-energies 
to these rates is chosen. 
From this procedure we find the relation
$E_{k}^0 = 3.0 \, {\rm meV} + 0.5 \, k_BT$ with the lattice temperature $T$, 
used for all simulations presented here. 
The second typical energy $E_{k'}^0= E_{k}^0+ \Delta E$ is then chosen to reflect the 
level difference $\Delta E$ as discussed in Ref.~\onlinecite{WackerIEEEJSelTopQuant2013}.

While the imaginary part of the remaining Green's function vanishes like $1/E_{k'}^2$ for large 
values, the real part has a logarithmic divergence, effectively making the self-energies
dependent on the $E_{k'}^{\rm max}$ chosen in the numerical implementation of Eq.~\eqref{EqSigma}.
To remedy this artefact we subtract the part of the integral over a certain
critical $E_k^{\rm cutoff} =  {\rm MAX}(E_{k}^0,E_{k'}^0)+M$ where $M$ is an appropriate 
margin. Using the non-interacting Green's function, we express the divergent part as
\begin{align} \nonumber
\Sigma_{\alpha\alpha}^{\rm div}(E) \approx & - X^{\rm elast}_{\alpha\alpha,\beta\beta}(E_{k}^0,E_{k'}^0) \\
\times & \log \left( \frac{E_{k'}^{\rm max}+E_\beta-E}{E_k^{\rm cutoff}+E_\beta-E} \right)
\end{align}
where we have restricted us to the diagonal parts of the scattering tensor $X^{\rm elast}$.
In order to remove the energy dependence we evaluate the right hand side using a typical
energy $E=E_\alpha+E_k$. In this work we use $M=20$ meV as this was found to give results
in good correspondence to fully momentum dependent calculations.

This provides us with a systematic procedure to evaluate and 
compensate for the artificial divergence in the real parts, and 
renders the self-energies independent on the integration limits in the
implementation, provided a sufficient range is used to cover all relevant 
physical processes.

\section{Choosing an effective electron temperature} \label{AppElectronTemp}

The electron temperature is by definition an eluding quantity when doing
non-equilibrium simulations. In any model where this thermodynamic 
intensive property is needed as an input parameter, the difficulties
will have to be circumvented in some way. While our standard model evaluates all distribution functions self-consistently and thus does not require this concept at all, we need the electron temperature for the plasmon occupations in the single plasmon-pole approximation used to approximate the GW result \cite{WingeJPhysConfSer2016}. 

In the following we will model the conduction band of the 
quantum cascade laser as one effective band, with an electron 
temperature $T_e$ as one of its properties.
A bias over this structure will heat the electrons, and they will subsequently 
relax emitting optical phonons. As the rate at which the electrons cool
increases with their temperature, a fixed point is reached. The energy balance
must thus fulfill
\begin{align} \label{EqBalance}
J\cdot(Fd) = en^{\rm 2D} E_{\rm LO} \left(\frac{1}{\langle \tau_{\rm em}\rangle }  
-  \frac{1}{\langle \tau_{\rm abs}\rangle}  \right)
\end{align}
where $J$ is the electron current density, $Fd$ the bias over one period, 
$n^{\rm 2D}$ the electron sheet density of one period, $E_{\rm LO}$ 
the energy of the longitudinal optical phonon and $(\tau_{\rm abs/em})^{-1}$ the rates of 
emitting or absorbing one such phonon, respectively. Here 
the scattering times have been averaged over a statistical distribution.
Acoustic phonon scattering is assumed to be small, 
and electron scattering only 
able to redistribute the carriers according to the electron 
temperature. 

If the rates can be expressed as functions of electron temperature, we can extract
this if the current and bias are known. For a bulk system the emission rate from a 
state with wave-vector $\mathbf{k}$
is given by 
\begin{align}
\Gamma_\mathbf{k} = \frac{C}{(2\pi)^2} \frac{\pi}{\alpha k} \Theta(E_k-E_{\rm LO}) \log \left| \frac{k+k_0}{k-k_0} \right| 
\end{align}
where we have summed over all possible final states. 
Here, $k_0=\sqrt{(E_k-E_{\rm LO})/\alpha}$ with $\alpha=\hbar^2/2m_{eff}m_e$ and $k=|\mathbf{k}|$. 
The constant $C$ is given by
\begin{align}
C = \frac{(n_{\rm LO}+1)}{\hbar}\frac{e^2 E_{\rm LO}}{2\epsilon_0} \left(\frac{1}{\epsilon(\infty)}-\frac{1}{\epsilon(0)} \right)
\end{align}
with the phonon occupation number $n_{\rm LO}=(1-\exp(E_{\rm LO}/k_{\rm B}T_L))^{-1}$, 
and the relative permittivities $\epsilon(0)$ and $\epsilon(\infty)$ at $E=0$ and infinity, 
respectively. 

Averaging over a 
Maxwell-Boltzmann distribution in 3D
we find
\begin{align} \nonumber
\frac{\sum_\mathbf{k} f_\mathbf{k} \Gamma_\mathbf{k}}{\sum_\mathbf{k} f_\mathbf{k} } 
&= \frac{2}{\sqrt{\pi}(k_{\rm B}T_e)^{3/2}} \\ 
&\times\int_{E_{\rm LO}}^\infty \d E_k \e^{-\frac{E_k}{k_{\rm B}T_e}} \frac{C}{4\pi\sqrt{\alpha}}  
\log \left| \frac{k+k_0}{k-k_0} \right| ,
\end{align}
which is now independent of $\mathbf{k}$ and a function only of electron temperature. 
In order to get analytical expressions the integrand can be linearized, and this yields the result
\begin{align} \nonumber
\frac{\sum_\mathbf{k} f_\mathbf{k} \Gamma_\mathbf{k}}{\sum_\mathbf{k} f_\mathbf{k} } 
 \approx 2\e^{-\frac{E_{\rm LO}}{k_{\rm B}T_e}} \frac{C}{4\pi}  \frac{1}{\sqrt {E_{\rm LO}\alpha}}.
\end{align}
for $\langle\tau_{\rm em}\rangle^{-1}$ and a similar expression is easily obtained for the absorption process. The scattering times
in Eq.~\eqref{EqBalance} are thus known and solving for electron temperature 
yields the final result as
\begin{align}
k_{\rm B}T_e  &= \frac{-E_{\rm LO}}{\log \left[ 2 \frac{J\cdot(Fd)}{n^{\rm 2D} C \sqrt{E_{\rm LO}\alpha} }+ \e^{-\frac{E_{\rm LO}}{k_{\rm B}T_L}} \right] }.
\end{align}
Here the low power limit can be seen as the electron temperature will approach the lattice temperature. As the electric
power increases the electron gas is heated. As an example, we find an electron temperature of 130~K for $Fd=50$~mV,
$J=1000$~A/cm$^2$, $n^{\rm 2D} = 3.0\cdot10^{10}/{\rm cm}^2$ and a lattice temperature of 77~K.

%\bibliographystyle{aipnum4-1}
%\bibliography{refs_QuantTrans,footnote}

\begin{thebibliography}{60}
\bibitem [{bac()}]{backtravel}%
  \BibitemOpen
  \href@noop {} {}\bibinfo {note} {A full calculation is difficult, as standing
  waves provide an inhomogeneous profile contrary to main simulation
  assumptions.}\BibitemShut {Stop}%
\end{thebibliography}


\end{document}